# Azimuthal modulation instability, breathers and solitons in ring-core optical fibers

B. Kibler[*] and P. Béjot

*Laboratoire Interdisciplinaire Carnot de Bourgogne,*

*UMR 6303 CNRS-Université de Bourgogne, 21078 Dijon, France*

[*]*bertrand.kibler@u-bourgogne.fr*

**Abstract:** We numerically investigate azimuthal modulation instability in an optical fiber supporting orbital angular momentum modes only, i.e. a vortex fiber, by means of the scalar multimode unidirectional pulse propagation equation. We demonstrate that the nonlinear stage of azimuthal modulation instability taking place in such a ring-core fiber, with anomalous rotation group-velocity dispersion, can be simply described by analytical breather solutions of the corresponding nonlinear Schrödinger equation. Azimuthal soliton dynamics and nonlinear compression are also unveiled as well as specific spatial rotating features as a function of the topological charge involved. Our results open a new route for studying transverse nonlinear waves in optical fibers and for manipulating orbital angular momentum states.

## 1   Introduction

Optical self-trapped beams induced by self-focusing instabilities have been studied since the late 1960s [1-4]. Typically, the self-focusing of a high-power laser beam in nonlinear dielectric medium has been shown to induce the azimuthal-symmetry breaking of a cylindrical Gaussian beam, thus giving rise to multiple filaments in space. Such a transverse instability corresponds to growth of perturbations which depend on the azimuthal angle in cylindrical coordinates. Periodic beam breakup as well as recurrence phenomena have also been unveiled in a saturable nonlinear medium [5]. More recently, the azimuthal symmetry-breaking instability of ring-profile vortex beams has been intensively investigated in several Kerr-like and quadratic media [6-14]. Nonlinear propagation of optical vortices still remains of high interest and challenging for optimizing the creation, manipulation and detection of structured light. Indeed, the class of self-trapped ring-like beams first introduced with a spiral phase structure [15], and later extended to necklace beams carrying or not orbital angular momentum (OAM) [16], has gained a lot of attention during the last two decades, since representing two-dimensional spatial solitons with numerous potential applications to photonics. Nowadays, one has to tackle even more complex light objects including the study of OAM-carrying beams and space-time wave packets which involve the simultaneous coupling of many degrees of freedom through the properties of the medium [17-20].

Combining OAM-carrying beams within fiber technology has represented a major step forward in space-division multiplexing for ultra-high capacity communications [21-22]. In particular, specific ring-core fibers have been developed to optimize radial light confinement, thus supporting the propagation of a large number of OAM modes without detrimental mode couplings for efficient data transmission. While in free-space systems and bulk media, an infinite and dense number of transversal modes can propagate, only a discrete and finite number of modes is supported by a given optical fiber. The modal distribution of multimode optical fibers then provides a discretization of fiber properties and related propagating space-time wave packets [19-20,23-24].

In this work, we theoretically and numerically investigate the particular case of nonlinear focusing wave propagation in an optical fiber supporting orbital angular momentum modes, i.e., a vortex fiber. We show that such a ring-core fiber exhibits anomalous discretized OAM-modal dispersion, thus favoring the spontaneous emergence of azimuthal modulation instability. The latter can be simply analyzed by the one-dimensional focusing nonlinear Schrödinger equation (NLSE) and related exact analytical breather solutions. We also demonstrate that azimuthal soliton dynamics may occur as well as spatial rotating features as a function of the topological charge involved.

## 2    Theory and numerical modeling

### 2.1    OAM modes properties in a vortex fiber

Several modal bases in cylindrical coordinates can describe the light propagation in optical fibers.

Fiber guided modes can be implemented with conventional linearly polarized (LP) modes, and they can be formed from linear combinations of vector eigenmodes of the fiber. LP mode basis results from a scalar approach of the wave equation, and designated usually as $LP_{lm}$ modes, where indices $l$ and $m$ are the azimuthal and radial indices. Only the $l = 0$ modes are circularly symmetric. The remaining modes exhibit rotational symmetry, where the field distribution is symmetric under rotations of $2\pi/l$. Optical fibers can also support OAM (higher-order) modes with helical phase front by correctly superposing orthogonal LP modes or vector modes, i.e. the even and odd modes for each $LP_{lm}$ or $HE_{l+1,m}$ and $EH_{l-1,m}$ modes with $\pm\pi/2$ phase shift in the linear combination. The resulting OAM modes are denoted as $OAM_{l,m}$ where $l$ and $m$ subscripts denote the azimuthal and radial indices, respectively. In the vector case, additional superscript is used to indicate the direction of the circular polarization. Here, $l$ is the topological number related to the phase front of the OAM mode, $m$ describes the number of nulls radially (rings) in the intensity profile of the OAM mode. As a simple example we can write: $OAM_{l=\pm 1, m=1} = (LP_{11}^{even} \pm iLP_{11}^{odd})/\sqrt{2}$.

In a weakly-guiding multimode fiber, we commonly make use of LP mode basis due to the similar propagation constants of vector eigenmodes and degeneracy. In general, LP mode supporting fiber can exhibit radial higher-order modes, where each mode field distribution is different. By contrast, OAM mode supporting fibers used for optical communications are designed to carry only radial first-order mode, by means of a depressed-inner region inside the core, which forms a core with ring shape. Such an index contrast also allows to manage the separation degree between vector modes and detrimental mode couplings during propagation [25]. In the following, we make use of such a fiber profile, more specifically the one developed in Ref. [26], with an air core and an annular index profile. Such a fiber has been experimentally validated to guide a large number of OAM modes, up to $l = 9$, with initial excitation from free-space coupling of OAM beams. In our study, we first neglect the vector nature of guided modes for better clarity, nevertheless it does somewhat impact our findings as discussed later in the article. Indeed, we specially used a ring-core fiber to decouple the azimuthal from the radial dependence of guided modes, since only the radial first-order mode can be guided and all remaining OAM modes that depend on $l$ exhibit very similar intensity patterns. In this regard, Fig. 1 depicts the refractive index fiber of the fiber chosen (subplot (a)), the corresponding effective refractive indices of scalar OAM guided modes (subplot (b)), as well as some examples of intensity and phase distributions of these modes (subplots (c-g)). Modal calculations were performed by means of the numerical method developed in Ref. [27]. We note that the fiber ring core made of GeO2-doped silica matches the donut shaped OAM fields by means of the high-index contrast provided by the centered air hole in the fiber. The width of the doped ring core was set to limit the guidance only to modes with $m = 1$ (i.e., having a single ring in their intensity profile) at the wavelength of 1.03 µm. The fiber supports a large number of OAM mode ($0 \leq |l| \leq 16$), such modes have an effective index that resides between the refractive index of the cladding and the maximum refractive index of the fiber. OAM modes with higher topological charge propagate faster than those with lower topological charge. Their intensity ring shape is very similar whatever the topological charge, as confirmed by their constant effective mode area (see Fig. 1g). In Fig. 1(h), we plotted rotation group-velocity dispersion (R-GVD) that OAM wave-packets may experience in our fiber, calculated from the (discrete) second derivative of the propagation constants with respect to $l$. We clearly note a nearly-constant anomalous dispersion (negative R-GVD) with topological charge, except for the latest guided modes ($|l| > 15$).

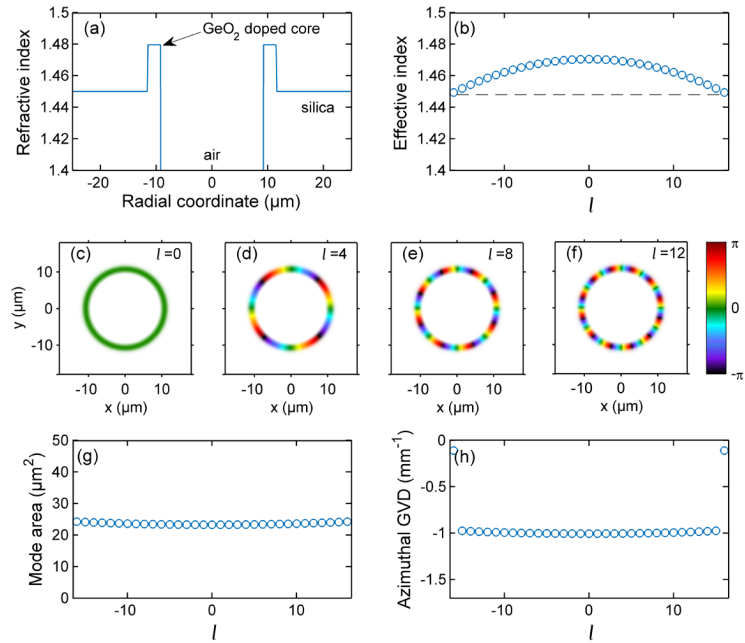

**Figure 1.** (a) Refractive index profile of the ring-core fiber under study given at 1.03 µm wavelength. (b) Corresponding calculated effective indices of OAM guided modes (dashed line indicates pure silica index). (c-f) Intensity and phase distribution of some OAM guided modes, namely $l = 0, 4, 8,$ and $12$, respectively. (g-h) Calculated effective mode area and rotation group-velocity dispersion as a function of the topological charge $l$.

## 2.2 Nonlinear propagation modeling

In the following we make use of a 3D+1 numerical approach of nonlinear wave propagation in OAM-carrying multimode fibers based on the multimode unidirectional pulse propagation equation (MM-UPPE) derived in Refs. [19,27], which describes the evolution of the complex electric field in the scalar approximation. In the context of fiber propagation, using a complex representation of the electric field $\xi$ (expressed so that $|\xi|^2 = I(r, \theta, t)$, $I$ being the pulse intensity), MM-UPPE writes as follows:

$$\partial_z \bar{\xi}(l, m, \omega) = iK_z\bar{\xi} + \frac{in_{eff_0}n_2\omega^2}{c^2 K_z}\overline{|\xi|^2\xi} \qquad (1)$$

where $\bar{\xi}(l, m, \omega)$ refers to the electric field in the modal basis, $K_z(l, m, \omega)$ is the propagation constant of the OAM mode $(l, m)$ at the frequency $\omega$, $n_{eff_0}$ is the effective refractive index of the fundamental mode at $\omega_0$, $n_2$ is the nonlinear refractive index of the fiber glass (here for silica glass, we used $n_2 = 3.2 \; 10^{-20}$ m²/W). In contrast to Ref. [19], we here neglect the Raman contribution into the nonlinear response, only the Kerr effect will be investigated. The impact of usual fiber losses (less than 10 dB/km) is not studied since being negligible over the considered propagation distance. All the numerical results shown below are obtained by solving the propagation equation (Eq. 1) without any additional approximation, and more particularly by means of a split-step modal algorithm as described in Ref. [27]. The present numerical model has been already tested successfully with experimental works on ultrashort pulse nonlinear propagation in MMFs [24].

In the limiting case of monochromatic wave propagation (i.e., continuous-wave pump $\omega = \omega_0$) in our vortex fiber, where only one radial mode is supported (i.e., $m = 1$) and the considered field whose OAM spectrum is centered around $l = l_0$, one can simplify Eq. (1) to derive suitable analytical tools and solutions. We expand the resulting $K_z(l, m = 1, \omega = \omega_0)$ in a Taylor series around $l_0$ (at $\omega_0$) in the linear term of Eq. (1), as $K_z(l) = K_0 + K_1(l - l_0) + K_2(l - l_0)^2/2$, where $K_p = \{\partial^p K_z/\partial l^p\}_{l=l_0}$. Physically speaking, the term $K_1$ (resp. $K_2$) corresponds to the rotation group velocity (resp. rotation group velocity dispersion) of the beam. In addition, we can simply refer to the first-order approximation of the frequency-dependent nonlinearity in Eq. (1), so that $n_{eff_0}n_2\omega^2/(c^2 K_z) \approx n_2\omega_0/c$. Moreover, without loss of generality, one can use a local frame propagating (resp. rotating) at the phase velocity $v_\varphi$ (at the rotation group velocity $K_1$) of the mode $l = l_0$. This can be made by the following change of variables: $t \to t - z/v_\varphi$, $\theta \to \theta - K_1 z$, $z \to$

$z$. Finally, as the radial part of the electric field $F(r)$ can be considered in very good approximation as independent of the topological charge, one can introduce the separation of variables by assuming the following form: $\xi(r,\theta,z) = F(r)A(\theta,z)e^{il_0\theta}$. We can then normalize $A$ such that $|A|^2$ represents the angularly-resolved optical power through $|A|^2 = P_0 = \int r\,|\xi|^2 dr / \int r\,|F(r)|^2 dr$. After these simplifications in Eq. (1), we obtain the following nonlinear Schrödinger (NLS) equation that describes the evolution of azimuthal envelope profile along the fiber:

$$\partial_z A(\theta,z) = i\frac{K_2}{2}\partial_\theta^2 A + \frac{in_2\omega_0}{cA_{\text{eff}\_\theta}}|A|^2 A \qquad (2)$$

where $A_{\text{eff}\_\theta} = [\int r\,|F(r)|^2 dr]^2 / \int r\,|F(r)|^4 dr$. One can also define the usual fiber nonlinear parameter as $\gamma = \frac{n_2\omega_0}{cA_{\text{eff}\_\theta}}$ (here $\gamma = 8.3$ W$^{-1}$ km$^{-1}$). On the whole, we simply recover a standard focusing NLS equation, thus allowing to investigate modulation instability, breather and solitons solutions along azimuthal coordinate $\theta$ instead of time coordinate for instance.

### 2.3 Azimuthal modulation instability and theoretical solutions

By using the above spatial NLS equation (Eq. 2) with anomalous R-GVD from our vortex fiber (i.e. focusing regime), one can easily derive the criterion for modulation instability (MI) found from linearized equations for small perturbations of the envelope profile of an OAM mode (i.e., linear stability analysis [28]). The corresponding azimuthal MI gain exists only for a limited number of azimuthal indices (i.e., topological charges) of modulation, namely for $|l| < l_C$, and is given by $g(l) = |K_2(l - l_0)|\sqrt{l_C^2 - (l - l_0)^2}$, where $l_C^2 = 4\gamma P_0 / |K_2|$. The gain spectrum is symmetric with respect to $l = l_0$, and it reaches its maximum at $l_{\max} = l_0 \pm l_C/\sqrt{2}$. Given that growing perturbations that can experience a net power gain are solely characterized by discrete azimuthal indices (integer numbers for $l$ values), it becomes evident that Akhmediev breather (AB) solutions provides the exact formulation of this azimuthal MI process in both induced and spontaneous regimes. The Akhmediev breather describes the full nonlinear evolution with z of a wave with initial constant amplitude on which is superimposed a small periodic perturbation taking here the form of a $\theta$-dependent modulation. This breather solution exhibits a single growth-return cycle featured by localization in the z direction. The AB solution to Eq. (2) can be written explicitly in the following form [29-30]:

$$A_{AB}(\theta, z) = \sqrt{P_0} \frac{(1-4a)\cosh[bz] + ib\sinh[bz] + \sqrt{2a}\cos[(l_{mod}-l_0)\theta]}{\sqrt{2a}\cos[(l_{mod}-l_0)\theta] - \cosh[bz]} \qquad (3)$$

The above equation represents the family of first-order AB solutions with a single independent parameter, namely the perturbation azimuthal index $l_{mod}$. The solution is valid over the range of modulation frequencies that experience MI gain. The coefficients $a$ and $b$ are defined as follows: $2a = 1 - [(l_{mod} - l_0)/l_C]^2$ and $b = [8a(1-2a)]^{1/2}$. The coefficient $a$ depends on input parameters and varies in the interval $0 < a < 1/2$, while the parameter $b > 0$ is directly linked to the MI growth. The maximum MI gain condition corresponds to $b = g/(2\gamma P_0) = 1$, and it occurs when $a = 1/4$, i.e. $l_{mod} = l_{max}$. The solution in Eq. (3) describes an evolving periodic train of spots onto a ring-profile vortex beam with azimuthal period $\theta_{mod} = 2\pi/(l_{mod} - l_0)$. The individual spots have maximum amplitude and minimum width at $z = 0$.

In this focusing regime where nonlinearity and dispersion counterbalance each other, it is obvious that standard NLS soliton solutions can be analyzed, as well as many other unstable solutions of Eq. (2) beyond the Akhmediev breather. We do not provide an exhaustive list of all potential solutions that can be investigated in our scalar approach of vortex fibers, we refer the reader to some optical studies of these complex nonlinear wave solutions performed in the temporal domain [31-34]. An easy transposition can be applied as described above from time to space coordinate. Here, we simply recall the fundamental soliton solution as $A_s(\theta, z) = \sqrt{P_s}\text{sech}(\theta/\Delta\theta)$, where the soliton width is $\Delta\theta = (|K_2|/\gamma P_s)^{1/2}$. Note that higher-order solitons can be studied in a simple way by considering that the peak power necessary to excite the $N$th-order soliton is $N^2$ times that required for the fundamental soliton [28].

## 3 Results

### 3.1 Azimuthal modulation instability

We first investigate numerically the spontaneous (noise-driven) regime of azimuthal MI by means of the full MM-UPPE modeling (see Eq. 1). We only assume a monochromatic wave propagation (i.e., continuous-wave pump $\omega = \omega_0$) in our vortex fiber, and we consider the fundamental OAM mode of the fiber as the input pump mode ($l_0 = 0$), with input power $P_0 = 100$ kW, together with a broadband white noise background. Figure 2(a-c) shows a typical single realization of our simulation for a 15.5 mm propagation distance. A localized structure is observed to spontaneously emerge along the azimuthal coordinate in Fig. 2a with its corresponding power profile depicted in Fig. 2b, this phenomenon is obtained after a strong development of the MI spectrum as shown in Fig. 2c.

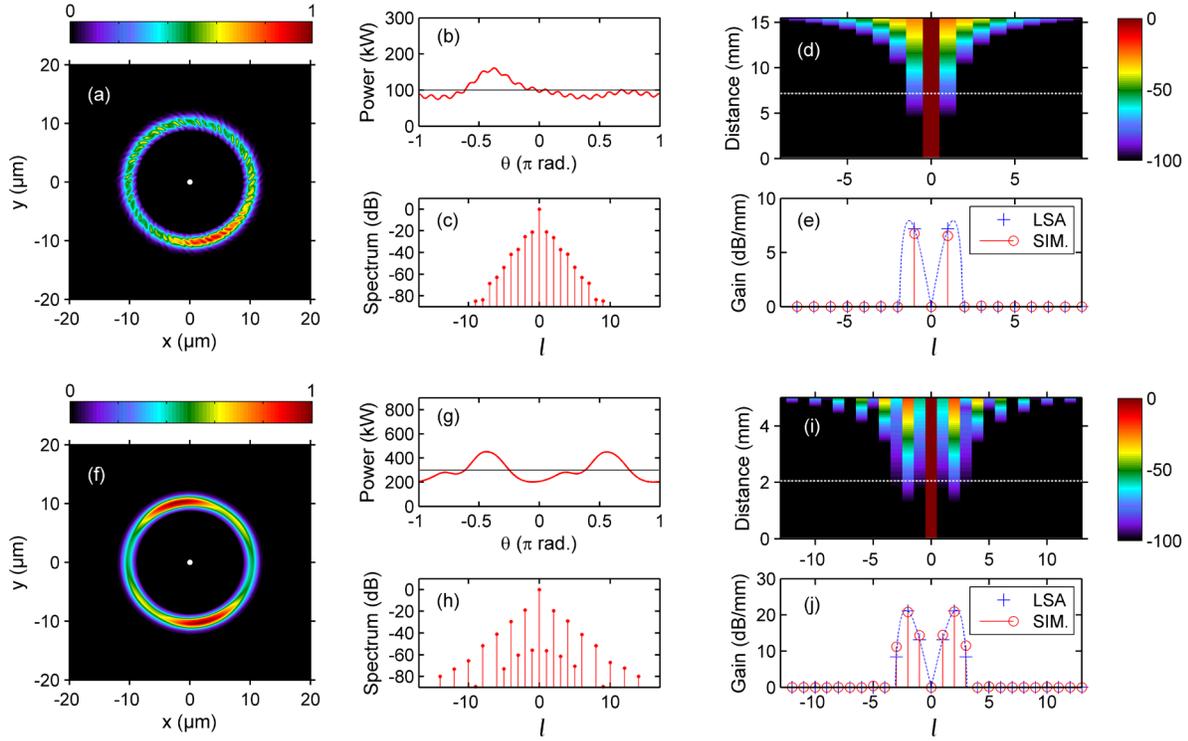

**Figure 2.** Spontaneous development of azimuthal modulation instability for distinct input powers. (a) Output normalized fluence distribution in space (single realization) after a propagation distance of 15.5 mm. Fundamental mode ($l = 0$) of the fiber is injected with input power $P_0 = 100$ kW. (b) Corresponding azimuthal power profile. (c) Corresponding power spectrum. (d) Spectral evolution (log. scale) as a function of propagation distance (averaged over 40 realizations). (e) Azimuthal MI gain calculated numerically over first 7 mm of propagation (see white dashed line in subplot (d)) compared to analytical theory from linear stability analysis (LSA). Dashed blue line is indicated for visualizing usual MI gain curve, here only one azimuthal index is amplified. (f-j) Similar results obtained for input power $P_0 = 300$ kW and 5-mm propagation distance. Azimuthal MI gain is here analyzed after 2 mm of propagation.

The evolution of the power spectrum as a function of propagation is reported in Fig. 2d, our simulation results were averaged over 40 realizations. We then compared the simulated MI gain in the first steps of propagation (linear stage of MI) with theoretical MI gain obtained from Eq. 2. An excellent agreement can be noticed in Fig. 2e, where only one azimuthal index ($l = 1$) satisfies the MI criterion.

As another example in Fig. 2(f-j), we analyze the results obtained for similar input conditions but using a higher power $P_0 = 300$ kW. In that case, two localized structures are emerging in the transverse coordinates, in particular on the finite background of the fundamental mode (see Fig. 2f). Corresponding power profiles in direct and Fourier spaces are depicted in Fig. 2(g-h). We easily note that the number of localized structures is related to the azimuthal index ($l = 2$) which dominates the fully developed MI spectrum and drives the subsequent cascaded harmonics. This is corroborated by the spectral evolution averaged over 40 simulations shown in Fig. 2(i-j), where the MI gain after 2 mm of propagation is well predicted by the linear stability analysis. As expected the higher input power increases the MI bandwidth, three one azimuthal indices ($l = 1,2,3$) experience significant gain, $l = 2$ being the maximally amplified topological charge with a gain value beyond 20 dB/mm. The above results confirm that azimuthal MI may spontaneously occur for OAM modes of a vortex fiber over very short propagation distances for high input powers. Moreover, the linear stage of MI dynamics can be well predicted by a linear stability analysis of the approximated model provided by Eq. (2). The azimuthal index that experienced maximum MI gain drives the number of modulations and subsequent localized structures emerging along the azimuthal coordinate in direct space. We also checked that we recover similar MI dynamics by considering an higher OAM mode ($l_0 \neq 0$) of the fiber as the input CW mode. The only difference is the rotation around the propagation axis of the emerging localized structured. The direction of rotation and its period along the propagation simply depends of the index and sign of input topological charge [19].

### 3.2 Akhmediev breathers and their interactions

In a second set of simulations, we study the seeded regime of azimuthal MI by means of the full MM-UPPE modeling. We again assume a CW pumping in the fundamental OAM mode ($l_0 = 0$) of the fiber, with input power $P_0 = 300$ kW, but now we superimpose a low-power CW seed in higher-order OAM modes $|l| = 3$. In this configuration, we excite an Akhmediev breather ($a = 0.03$) with an approximate sinusoidal perturbation along the azimuthal coordinate. Corresponding numerical results are shown in Fig. 3(a-f). A growth-decay cycle is clearly observed with 10-mm-long propagation distance in both direct and Fourier spaces (see Fig. 3(a-b)).

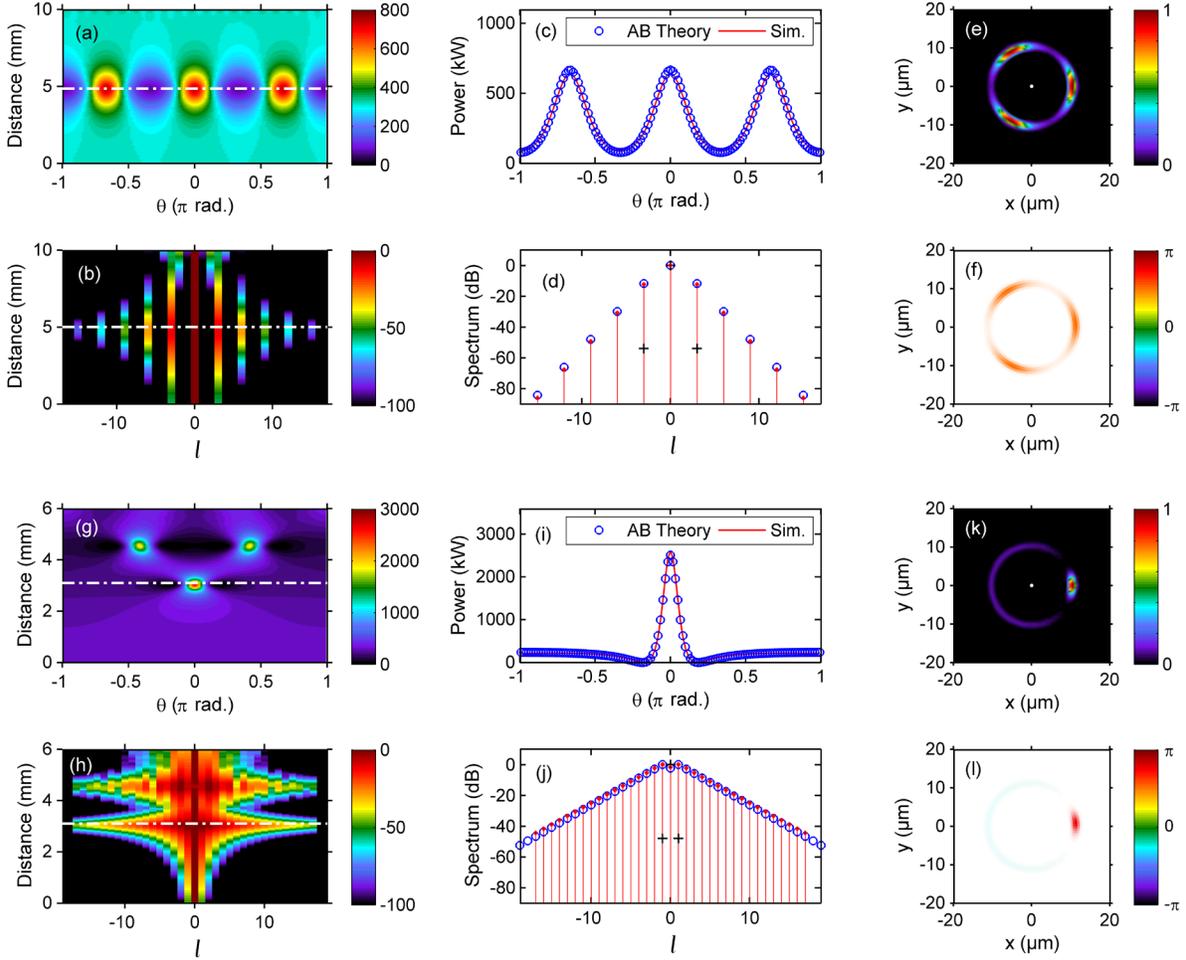

**Figure 3.** Non-ideal excitation of Akhmediev breathers ($a = 0.03$ and $a = 0.45$) in the regime of seeded azimuthal modulation instability. (a-b) Evolution of spatial and spectral power profiles along propagation distance for $a = 0.03$ (seeding at $|l| = 3$), respectively. (c-d) Corresponding simulated profiles at maximum breather compression (see white dashes lines in subplots (a-b)), namely for a distance of 4.8 mm, compared with AB theory from Eq. (3). Input spectrum is indicated with black crosses in subplot (d). (e-f) Output normalized fluence and phase distributions in space at maximal compression, respectively. (g-l) Similar results obtained for $a = 0.45$ (seeding at $|l| = 1$). Profiles at maximum breather compression are analyzed here at a distance of 3 mm.

At maximum amplification of the periodic perturbation (after a distance of 4.8 mm), power profiles in both spaces (Fig. 3(c-d)) are described in good agreement by the corresponding AB theoretical solution given by Eq. (3). The corresponding fluence and phase distributions in $(x, y)$ space (i.e., the fiber cross section) is depicted in Fig. 3(e-f). The constant phase along the azimuthal modulated profile agrees with the AB theory when $a < 0.125$. One can note that at the end of breather decay, we do not exactly recover the CW background due to non-ideal input excitation [30]. By pumping a higher azimuthal OAM mode instead of the fundamental one, one would observe the rotation of the breather in $(x, y)$ space with propagation (i.e., non-zero rotation group velocity).

As another example in Fig. 3(g-l), we analyze the results obtained for similar input conditions but using a low-power seed at $|l| = 1$. In that case, the governing parameter of excited Akhmediev breather is $a = 0.45$, close to the Peregrine breather limit ($a = 0.5$), thus explaining the stronger localization phenomenon in space of the single beam spot and its amplification factor close to 9 in terms of peak power with respect to the initial background [35]. It is worth noting that the strong spatial nonlinear focusing into a single spot over a small CW background (see Fig. 3(i,k)) is associated with dramatic spectral broadening over all OAM mode components of the fiber shown in Fig. 3j. Note that the AB theory does not take into account the limited number of OAM guided modes in the fiber under study, nevertheless the spectral truncation occurring in our simulation does not prevent the complete growth of the breather. Moreover, the profile at maximal compression in Fig. 3(i) exhibits two zero-intensity points, the real field experiences a sign inversion between the central peak and the continuous wave background as predicted by the theoretical solution. This corresponds to a $\pi$ phase shift between the two azimuthal parts of the beam as it can be seen in panel Fig. 3(l) representing the phase distribution (see light blue color for the CW background, and red color for the central spot). Another important phenomenon has to be pointed out in Fig. 3(g-h), namely the observation of higher-order modulation instability [36]. This instability arises from the nonlinear superposition of elementary instabilities (i.e., multiple breathers) associated with initial (non-ideal) single breather excitation, since harmonics of the initial perturbation also fall into the MI criterion. A second breather linked to $|l| = 2$ typically emerges at longer propagation distance (during the decay of the first breather), here its maximum amplification occurs at 4.55 mm. The evolution cannot be considered as an independent superposition of breathers, thus affecting here the dynamics observed in $(\theta, z)$ space. Note that the full pattern can be described analytically by using the Darboux transformation [36].

The nonlinear superpositions of breathers may be synchronized in $(\theta, z)$ space so that their intensity peaks collide at the same spatial location, giving rise to a higher-order breather solution characterized by an extreme peak intensity. Figure 4 shows the numerical results corresponding to the collision of two ABs, when assuming again a CW pumping in the fundamental OAM mode ($l = 0$) of the fiber, with input power $P_0 = 300$ kW, and the seeding of the higher-order OAM modes $l = +3$ and $l = -2$. The AB parameters under study are $a_1 = 0.03$ and $a_2 = 0.29$. Specific initial conditions by controlling the phase and seeding amplitude of the breathers are required for their efficient collision [32]. With opposite sign of the seeding topological charges, breathers experience inclined trajectory in $(\theta, z)$ space (distinct mean group velocity of AB compared to the previous symmetric perturbation studied in Fig. 3). The control of this group velocity difference between ABs favors their collision on a short propagation distance. Moreover, the intrinsic discretization of modal parameters makes here that any seeded MI dynamics will correspond to commensurate frequencies of ABs. The resulting wave is then periodic as shown in Fig. 4(a,c). Here, we see clearly that breather collision produces a high peak that is more than 9 times the initial average power after a distance of 2.9 mm, and associated with an extreme spectral broadening. An extremely localized spot is recovered in $(x, y)$ space (see Fig. 4(e)) with a secondary spot with same spatial phase (see Fig. 4(f)). The full wave profile is fully described by a second-order AB solution as reported in Fig. 4(c-d). We refer the reader to Refs. [29,32] for the complex analytical form of this solution.

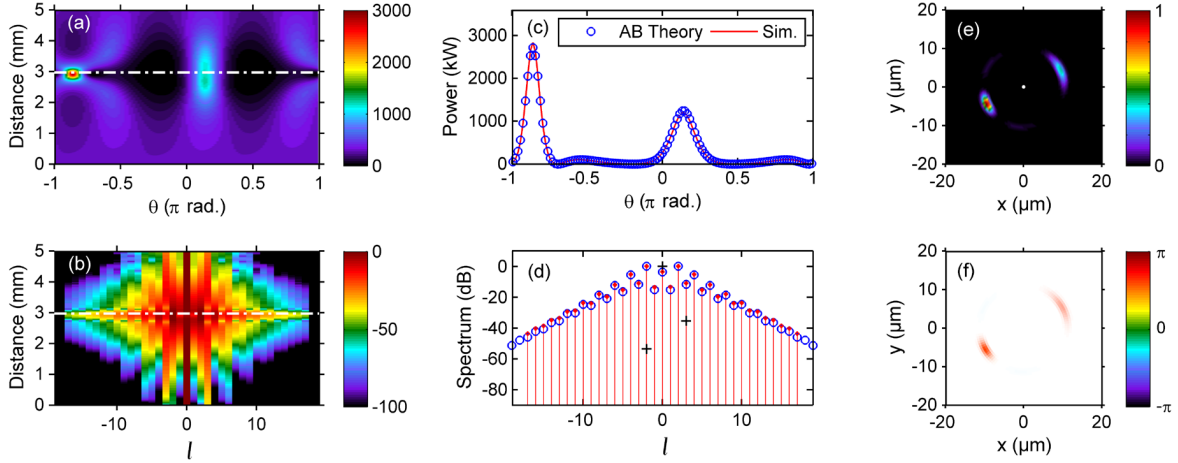

**Figure 4.** Synchronized collision between two Akhmediev breathers ($a_1 = 0.03$ and $a_2 = 0.29$) in the regime of seeded azimuthal modulation instability. (a-b) Evolution of spatial and spectral power profiles along propagation distance, respectively. (c-d) Corresponding simulated profiles at collision point (see white dashes lines in subplots (a-b)), namely for a distance of 2.9 mm, compared with analytical second-order AB solution of Eq. (3). Input spectrum is indicated with black crosses in subplot (d). (e-f) Output normalized fluence and phase distributions in space at collision point, respectively.

### 3.3 Solitons on zero background and nonlinear focusing

In a third set of simulations, we examine the existence of common soliton dynamics [28], here in $(\theta, z)$ space by means of the full MM-UPPE modeling. To this end, we verify the propagation from nearly exact input conditions corresponding to NLS soliton solutions. For simplicity, we assume a spectrally-shaped input wave for $|l| < 10$ constructed from the fundamental soliton written in section 2.3. Figure 5(a-f) shows the numerical results obtained for the fundamental soliton ($N = 1$), whereas the more complex dynamics of second-order soliton ($N = 2$) is depicted in Fig. 5(g-l). In both cases, there is an excellent agreement between numerical simulations and analytical solutions, even in the case of the extreme broadening beyond guided OAM modes depicted in Fig. 5(j). We also recovered typical phase features of NLS soliton solutions shown in Fig. 5(f,l).

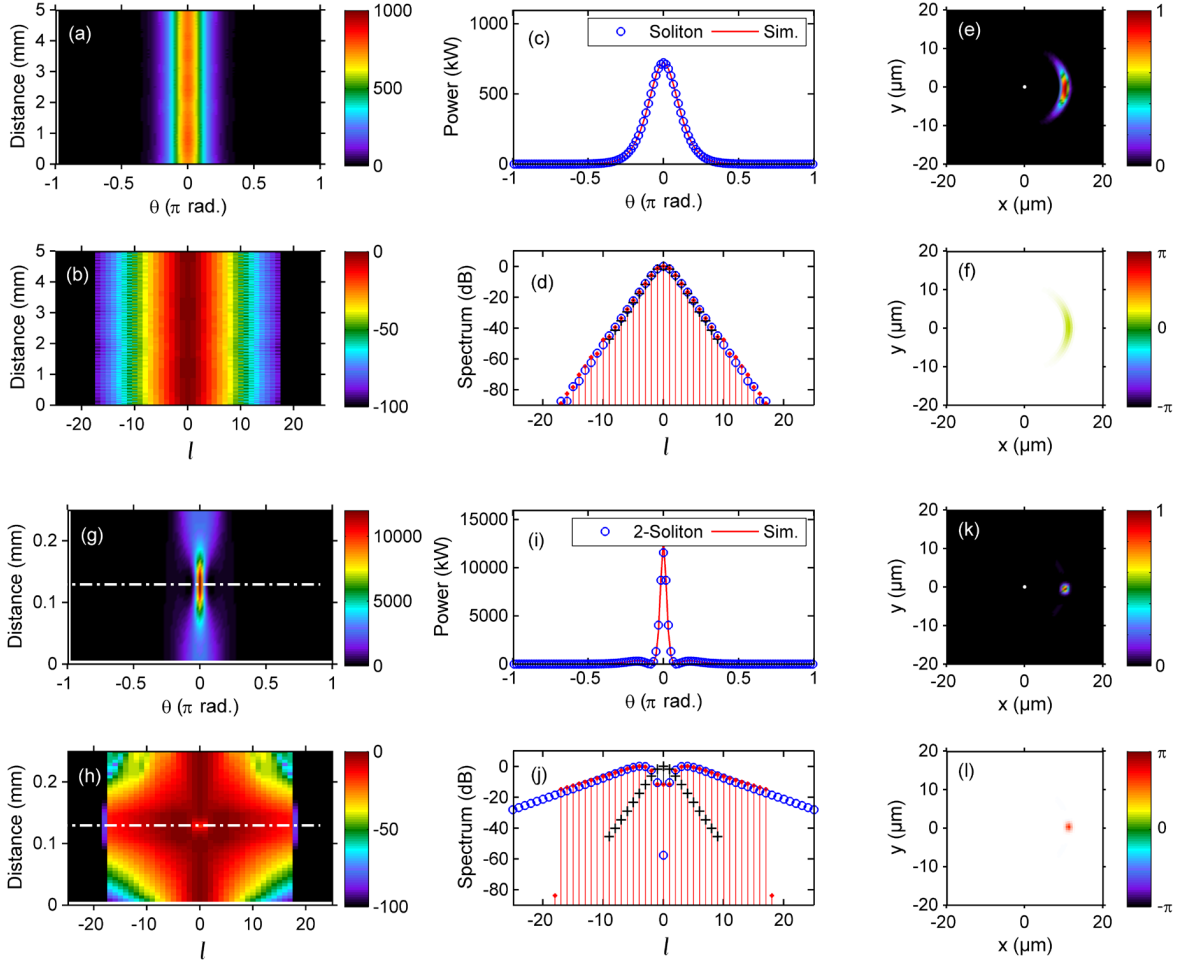

**Figure 5.** Azimuthal soliton dynamics in our vortex fiber. (a-b) Evolution of spatial and spectral power profiles along propagation distance for $N = 1$, respectively. (c-d) Corresponding simulated profiles after a distance of 5 mm, compared with analytical fundamental soliton solution of Eq. (3). Input spectrum is indicated with black crosses in subplot (d). (e-f) Output normalized fluence and phase distributions in space, respectively. (g-l) Similar results obtained for $N = 2$. Profiles at maximum soliton compression are analyzed here at a distance of 0.13 mm and compared with the corresponding analytical solution [28].

Besides nonlinear spatial compression in the framework of ideal soliton dynamics, one can simply consider the multiple four-wave mixing of an initial dual-OAM beat signal [37]. We considered a simple dual-OAM modal pumping ($l = 0$ and $l = +1$) of the fiber, and with an input average power equal to 51 kW. This corresponds to one period of a sinusoidal input wave over the azimuthal coordinate. In the focusing regime of NLS equation, this configuration is well understood since multiple four-wave mixing processes between the two CW pumps spontaneously occur in the Fourier domain without any power threshold [38]. High-quality compressed wave packets with a Gaussian shape and nearly constant phase, and without residual pedestals, can be obtained at maximal compression distance, as shown in Fig. 6. Empirical relations giving the optimal fiber length and average input power as a function of the azimuthal index difference $\Delta l$ of the source and the fiber dispersion were adapted from previous works in the time domain (see Ref. [39]). Here we obtained the following simulation parameters: $L_{opt} = 4\pi^2/(14K_2\Delta l^2) = 2.7$ mm, and $P_{opt} = 16K_2\Delta l^2/4\gamma\pi^2 = 51$ kW, respectively. By increasing the index difference between the topological charges of the CW pumps involved, one can also generate pedestal-free periodic spots along the azimuthal coordinate. It is also worth to note the continuous rotation of the OAM wave packet along the azimuthal coordinate due to the initial asymmetric pumping with respect to $l = 0$. The period of rotation is equal to $2\pi/|K_z(l=1) - K_z(l=0)| = 12$ mm. Note here the small rotation period observed in this ring-core fiber compared to standard step-index MMF [19], due to the large separation of propagation constants.

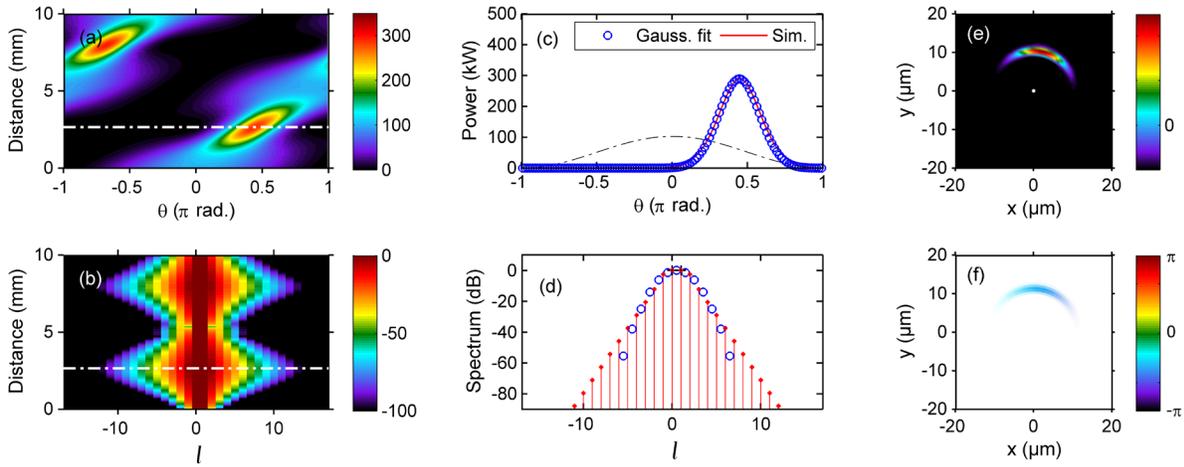

**Figure 6.** Azimuthal nonlinear compression from a simple dual-OAM beat signal. (a-b) Evolution of spatial and spectral power profiles along propagation distance, respectively. (c-d) Corresponding simulated profiles at maximal compression distance, namely at 2.7 mm, and compared with a Gaussian fit. Input wave profile and spectrum are indicated with black dashed line and black crosses in subplots (c) and (d), respectively. (e-f) Output normalized fluence and phase distributions in space at maximal compression, respectively.

## 4 Discussion

Our above results confirm that well-known nonlinear wave dynamics described by a 1D+1 spatial NLS equation can be easily retrieved in the transverse plane of a vortex ring-core fiber by means of OAM modes. Azimuthal MI was investigated in both spontaneous and seeded regimes. Non-ideal excitations or exact shaping of NLS solutions were also investigated successfully. In some cases, simple approximate input waves were specifically used for practical reasons, in terms of shaping and superposition of OAM modes, to encourage future experimental works. Different kinds of 2D-spatial shaping techniques could be used such as simple vortex plates but also 2D spatial light modulators to tailor and manipulate input OAM modes [25-26]. One specific requirement to observe the above transverse nonlinear dynamics was the use of high-power continuous or quasi-continuous waves (beyond tens of kW), in particular with the silica-based fiber under consideration. This power range is slightly above that of recent (time-domain) MI experiments in multimode gradient-index fibers [40]. In practice, most of previous MI experiments (even in single-mode fibers) made use of nanosecond pulse lasers that deliver high peak power and long enough pulses to facilitate the study of MI processes with various optical fibers and parameter regimes. Here, one could also consider some compact passively Q-switched solid-state lasers or ytterbium fiber lasers in the nanosecond pulse regime that produce peak powers of hundreds of kW in the 1-µm waveband. Therefore, particular attention might be paid to the fiber damage threshold.

For simplicity, we neglected the vector nature of guided OAM modes in the ring-core fiber under study (see Fig. 1) to provide the simplest description of nonlinear dynamics. However, such a fiber is commonly developed for allowing the propagation of OAM-carrying modes with extremely low cross-talks over long distances, since they lift degeneracy sufficiently to enable mode coupling free propagation [25,41]. An exact vector modal analysis of the ring-core fiber would divide the guided modes into two distinct groups of modes with orthogonal (spatially-dependent) polarization patterns. Each mode within a group is characterized by a distinct *total angular momentum* $j_l$, as the sum of the spin ($s$) and angular ($l$) momenta, and involves both circular polarization contributions $|\sigma_\pm\rangle$ (with opposite spin index $s = \pm 1$) [42]. The corresponding vector modes of both groups can be written as follows [42]:

$$|\Phi_{\pm,j_l}\rangle = F(r)e^{i(j_l-1)\theta} \begin{pmatrix} \cos\alpha_{j_l} & e^{2i\theta}\sin\alpha_{j_l} \\ \sin\alpha_{j_l} & -e^{2i\theta}\cos\alpha_{j_l} \end{pmatrix} \begin{pmatrix} |\sigma_+\rangle \\ |\sigma_-\rangle \end{pmatrix} \quad (4)$$

where $\alpha_{j_l}$ accounts for the relative amplitude contribution of the two circular polarization components for modes embedding a total angular momentum $j_l$. As an example, here the fundamental vector mode $|\Phi_{+,j_l=0}\rangle$ of the ring-core fiber is characterized by $j_l=0$ and azimuthally polarized (i.e., TE mode), it then involves two OAM values: $l = j_l - 1$ with $|\sigma_+\rangle$, and $l = j_l - 1 + 2$ with $|\sigma_-\rangle$. We refer the reader to Ref. [42] for a deeper vector modal analysis of such a ring-core fiber. The main features can be summarized as follows: (i) vector modes of the same group but with opposite $j_l$ index have exactly the same effective refractive index, (ii) all modes almost share the same radial power distribution as in the scalar approach, and (iii) both vector and scalar approaches give a similar anomalous rotation group-velocity dispersion, thus giving rise to a focusing nonlinear propagation regime. As a consequence, similar modulation instability and breather-like dynamics were also observed numerically by means of a multimode vector nonlinear propagation solver. Besides, it is worth to mention that the polarization of the vector modes gradually tends to a pure circular polarization as the $j_l$ value increases, thus reducing the nonlinear vector couplings.

All this corroborate our approximate scalar analysis provided in preceding sections since the signatures of scalar nonlinear dynamics might be somehow recovered beyond the intrinsic vector couplings. It is always an asset to have a scalar approach for providing some important physical insights before the development of future experimental works. Obviously, an accurate vector nonlinear modeling approach will be necessary for giving a complete physical picture as the one introduced recently in Ref. [42], but it would involve more complex theoretical developments than those employed here with the integrable NLS equation.

## 5 Conclusion

In summary, we studied the existence of transverse nonlinear dynamics in an optical fiber supporting orbital angular momentum modes only, i.e. a vortex ring-core fiber, by means of the scalar multimode unidirectional pulse propagation equation. In particular, we demonstrated that azimuthal modulation instability can take place in such a fiber, with anomalous rotation group-velocity dispersion, and it can be described by analytical breather solutions of the corresponding nonlinear Schrödinger equation. Vortex soliton dynamics, nonlinear beam compression, and four-wave mixing processes between OAM modes were also confirmed as possible nonlinear phenomena in such a fiber. The possibility of making use of an integrable 1-D NLS equation of studying 2D nonlinear waves offers a multitude of further developments based on the whole family of NLS solutions. Fundamental instabilities and recurrent oscillations within the nonlinear propagation of OAM modes could be investigated [43]. Our results also open a new route for studying transverse nonlinear waves in optical fibers, but it also provides an interesting alternative to current fiber solutions [44] in order to manipulate OAM states and signals, in particular by means of nonlinear OAM converters and amplifiers through the above-studied parametric process.

**Acknowledgments.** This work was supported by the Agence Nationale de la Recherche (EIPHI Graduate School, Contract No. ANR-17-EURE-0002, HELIX project) and the Bourgogne Franche-Comté Region.

**Notes.** The authors declare no competing financial interest.